\documentclass[fleqn,12pt]{article}
\usepackage[affil-it]{authblk}
\usepackage[english]{babel}
\usepackage{blindtext}

\providecommand{\keywords}[1]{\textbf{\textit{Keywords:}} #1}
\providecommand{\msc}[1]{\textbf{\textit{2010 MSC:}} #1}

\usepackage{graphicx}
\usepackage{amsmath}
\usepackage{amsthm}
\usepackage{amssymb}
\usepackage{amsfonts}
\usepackage[hidelinks]{hyperref}
\usepackage{booktabs}

\usepackage{longtable}
\usepackage{cite}
\usepackage{mathrsfs}

\usepackage[title,titletoc,toc]{appendix}
\newtheorem{thm}{Theorem}[section]

\newtheorem{rem}{Remark}

\newtheorem{case}{Case}
\newtheorem{subcase}{Subcase}
\newtheorem{prof}{Proof}
\usepackage{chngcntr}
\counterwithin{case}{section}
\counterwithin{case}{subsection}
\counterwithin{subcase}{case}
\counterwithin{exmp}{section}
\counterwithin{rem}{section}
\counterwithin{red}{section}
\counterwithin{prof}{section}

\usepackage[margin=1in,left=1in,right=1in]{geometry}
\usepackage{subcaption}
\usepackage{float}
\usepackage{placeins}
\allowdisplaybreaks
\usepackage{appendix}
\title{Infinite dimensional symmetry group, Kac-Moody-Virasoro algebras and integrability of  Kac-Wakimoto equation}

\author[1]{Manjit Singh\thanks{corresponding author: manjitcsir@gmail.com}}

\affil[1]{%
    Yadavindra College of Engineering Punjabi University Guru Kashi Campus Talwandi Sabo--151302, Punjab, India.}

\begin{document}

\maketitle
\begin{abstract}
An eighth-order equation in (3+1)-dimension is studied for its integrability. Its symmetry group is shown to be infinite-dimensional and is checked for Virasoro like structure. The equation is shown not to have Painlev$\acute{\rm e}$ property. One and two-dimensional classifications of infinite-dimensional symmetry algebra is also given. 
\end{abstract}
\keywords{Lie symmetries, Kac-Wakimoto equation, Virasoro like algebra, Integrability.}\\
\msc{70H07, 17B65, 17B67, 17B68.}\\
\section{Introduction}
Over the last three decades, there have been various approaches for solving nonlinear partial differential equations, including computational and analytical methods. Above all the available methodologies, the Lie group method continues to prove its competence. The Lie group approach not only helps to solve nonlinear equations for solutions that are physically important, but also helps to detect the inherent geometric properties of the equation \cite{ovsi,bluman1969general,olverbook,anco}. The symmetries of nonlinear partial differential equations, particularly infinite symmetries, play a vital role in the study of the integrability of the equation, especially when the equation admits Vrasoro like algebra. It has been seen for many integrable equations admitting Virasoro algebra, such as; Nizhnik-Novikov-Veselov equation \cite{novikov1986two}, nonlinear Schr$\Ddot{\mathrm{o}}$dinger \cite{fokas1994simplest}, sine-Gordon equation, (2+1)-dimensional long dispersive wave equation \cite{chakravarty1995some}, and other non-integrable  equations such as; Infeld-Rowlands equation \cite{faucher1993symmetry}, dispersive long-wave
equation \cite{paquin1990group} do not admit Virasoro alegebra. This coincidence of Virasoro like algebra may be helpful in investigating whether or not the equation could be integrable, though we need to investigate additional integrability parameters (like Painlev$\acute{\rm e}$ property)  to support the claim. So, in this work, we plan to investigate the integrability of the following eighth-order (3+1)-dimensional Kac-Wakimoto equation.:
\begin{align}
u_{8x}&+28\,u_{x}u_{6x}+28\,u_{xx}u_{5x}+70\,u_{xxx}u_{4x}+210\,u_{x}^{2}u_{4x}+420\,u_{x}u_{xx}u_{xxx}\nonumber\\
\label{KW:1}&+420\,u_{x}^{3}u_{xx}+a\,(u_{xxxy}+3\,u_{y}u_{xx}+3\,u_{x}u_{xy})+b\,u_{zz}+c\,u_{xt}=0,
\end{align}
where $a=-280\sqrt{6}, b=210, c=-240\sqrt{2}$. This equation is associated with affine Lie algebra $\mathfrak{e}_{6}^{(1)}$  was once derived by Kac and Wakimoto \cite{kac1989exceptional} in Hirota's bilinear form. Dodd \cite{dodd2008integrable} has obtained its one and two-soliton solutions which subsequently were corrected by Pekcan \cite{pekcan2016kac} and the author proved that the Kac-Wakimoto equation is not integrable in Hirota's sense. The non-integrability of the Kac-Wakimoto equation is further reinforced by Sakovich \cite{sakovich2016integrability}, and the author has shown that the equation does not possess the Painlev$\acute{\rm e}$ property. Lately Wang et. al. \cite{wang2020some}, have constructed the rational, kink-type breather and degenerate three-solitary wave solutions  for the equation \eqref{KW:1}. In this work, we mainly focus on establishing non-integrability of Kac-Wakimoto equation on the basis of existence of Virasoro like Lie algebra, and along with this we shall provide exclusive classification of infinite-dimensional Lie algebra of equation \eqref{KW:1} in one and two-dimension. 
\section{Lie group analysis}
In this section, we start with brief and relevant discussion on Lie group analysis. Consider a partial differential equations $F(x,y,z,t,u,\partial u)=0$ and one-parameter Lie group of point transformations  in the following the form:
\begin{align*}
&x^{*}=T_{1}(x,y,z,t;\epsilon), y^{*}=T_{2}(x,y,z,t;\epsilon),\\
&z^{*}=T_{3}(x,y,z,t;\epsilon), t^{*}=T_{4}(x,y,z,t;\epsilon),
u^{*}=U(x,y,z,t;\epsilon).
\end{align*}
 The differential equation $F$ shall be invariant under these one-parameter transformations if and only if
\begin{align*}
    F(x^{*}, y^{*}, z^{*}, t^{*}, u^{*},\partial u^{*})=F(x, y, z, t, u,\partial u),
\end{align*}
or more precisely
\begin{align*}
   \lim_{\epsilon\to 0}\frac{F(x^{*}, y^{*}, z^{*}, t^{*}, u^{*},\partial u)-F(x, y, z, t, u,\partial u^{*})}{\epsilon}=0.
\end{align*}
Above is nothing but Lie derivative of differential equation $F(x,y,z,t,u,\partial u)=0$ along the following direction:
\begin{align*}
    X=\left(\frac{\partial x^{*}}{\partial\epsilon}\right)_{\epsilon=0}\frac{\partial}{\partial x}+\left(\frac{\partial y^{*}}{\partial\epsilon}\right)_{\epsilon=0}\frac{\partial}{\partial y}+\left(\frac{\partial z^{*}}{\partial\epsilon}\right)_{\epsilon=0}\frac{\partial}{\partial z}+\left(\frac{\partial t^{*}}{\partial\epsilon}\right)_{\epsilon=0}\frac{\partial}{\partial t}+\left(\frac{\partial u^{*}}{\partial\epsilon}\right)_{\epsilon=0}\frac{\partial}{\partial u},
\end{align*}
or equivalently
\begin{align*}
     X=\xi_{1}\frac{\partial}{\partial x}+\xi_{2}\frac{\partial}{\partial y}+\xi_{3}\frac{\partial}{\partial z}+\xi_{4}\frac{\partial}{\partial t}+\eta\frac{\partial}{\partial u}.
\end{align*}
The eighth-order prolongation $X^{(8)}$ of above vector field when acts on the equation \eqref{KW:1} shall provide following set of infinitesimals:
\begin{eqnarray}{\label{KW:2}}
\begin{aligned}
    \eta=&-\frac{1}{6}\,yzf_{1}^{\prime\prime}(t)+\frac{1}{3}\,yf_{2}^{\prime}(t)+zf_{3}(t)+f_{4}(t),\\
    \xi_{1}=&-\frac{1}{2}\,zf_{1}^{\prime}(t)+f_{2}(t),
    \,\xi_{2}=c_{2},
    \,\xi_{3}=f_{1}(t),
    \,\xi_{4}=c_{1},
\end{aligned}
\end{eqnarray}
and consequently, six dimensional Lie algebra $\mathfrak{g}$ is obtained:
\begin{eqnarray}{\label{KW:3}}
\begin{aligned}
   X_{1}=&\frac{\partial}{\partial t},\,X_{2}=\frac{\partial}{\partial y},\\
   X_{3}=&-\frac{1}{2}zf_{1}^{\prime}(t)\,\frac{\partial}{\partial x}+f_{1}(t)\,\frac{\partial}{\partial z}-\frac{1}{6}yzf_{1}^{\prime\prime}(t)\,\frac{\partial}{\partial u},\\
   X_{4}=&f_{2}(t)\frac{\partial}{\partial x}+\frac{1}{3}yf_{2}^{\prime}(t)\,\frac{\partial}{\partial u}, X_{5}=zf_{3}(t)\,\frac{\partial}{\partial u}, X_{6}=f_{4}(t)\,\frac{\partial}{\partial u}.    
\end{aligned}
\end{eqnarray}
The Lie algebra $\mathfrak{g}$ is closed under Lie bracket $[X_{i}, X_{j}]=X_{i}X_{j}-X_{j}X_{i}$, which is commutator of two generators $X_{i}$ and $X_{j}$. The Lie algebra \eqref{KW:3} might not appear closed under Jacobi identity, but the closeness under Jacobi's identity can be achieved by change of basis. The detailed results of all the Lie commutators are listed in the \autoref{KW_Table:1}. 
\begin{table}[!ht]
\centering
    \begin{tabular}{c|cccccc}
\hline\hline\\
  $[X_{i}, X_{j}]$ & $X_{1}$ & $X_{2}$ & $X_{3}$ & $X_{4}$ & $X_{5}$ & $X_{6}$ \\ \\\hline\hline\\
  $X_{1}$&0&0&$X_{3}(f_{1}^{\prime})$&$X_{4}(f_{2}^{\prime})$&$X_{5}(f_{3}^{\prime})$&$X_{6}(f_{4}^{\prime})$\\[1ex]
  $X_{2}$&0&0&$-X_{5}\left(\frac{1}{6}f_{1}^{\prime\prime}\right)$&$X_{6}\left(\frac{1}{3}f_{2}^{\prime}\right)$&0&0\\[1ex]
  $X_{3}$&$-X_{3}(f_{1}^{\prime})$&$X_{5}\left(\frac{1}{6}f_{1}^{\prime\prime}\right)$&0&0&$X_{6}(f_{1}f_{3})$&0\\[1ex]
  $X_{4}$&$-X_{4}(f_{2}^{\prime})$&$-X_{6}\left(\frac{1}{3}f_{2}^{\prime}\right)$&0&0&0&0\\[1ex]
  $X_{5}$&$-X_{5}(f_{3}^{\prime})$&0&$-X_{6}(f_{1}f_{3})$&0&0&0\\[1ex]
  $X_{6}$&$-X_{6}(f_{4}^{\prime})$&0&0&0&0&0\\[1ex]
  \hline\hline
\end{tabular}
\caption{Commutation relations for Lie algebra \eqref{KW:3}}
\label{KW_Table:1}
\end{table}
\begin{rem}{\label{KW_RM:1}}
\normalfont The Virasoro algebra has basis consisting of generators $X_{m},\,m\in \mathbb{Z}$ satisfying Lie commutation in following the manner:
\begin{align*}
    [X_{m}, X_{n}]=(m-n)X_{m+n}+\frac{c}{12}(m^{2}-m)\delta_{m+n,0},
\end{align*}
where $c$ is a central element commuting with all the generators. In classical sense, $c=0$, but it plays crucial role in quantum mechanics  (for more details, see the ref. \cite{goddard1986kac,gungor2006virasoro,kac2013bombay}).  The presence of Virasoro algebra is a good predictor of integrability that can be seen in the typically integrable equations in (2+1)-dimension \cite{david1985subalgebras,champagne1988infinite,faucher1993symmetry,paquin1990group}. On the basis of above argument for Virasoro algebra, we can infer from the Lie commutations  written in the \autoref{KW_Table:1} that the Lie algebra $\mathfrak{g}$ does not contain Virasoro algebra. This supports the argument presented in the work \cite{pekcan2016kac} that Kac-Wakimoto equation is not integrable. Whereas the investigation of Zakharov–Strachan equation \cite{senthil1998lie} reveals that it do not admit Virasoro algebra like structure but it is still integrable. However, the integrability of an equation can be judged from number of arguments in favour of or against it. 
\end{rem}
\section{Classification of Lie algebra under adjoint transformation}
Each infinitesimal generator in the Lie algebra \eqref{KW:3} is capable of generating Lie group of point transformations \cite{olverbook,coggeshall1992group} through exponentiation
\begin{align}
\label{KW:4}\tilde{\psi} = \exp\left(a_{i}X_{i}\right)\psi,\;i = 1\dots 6,\; \text{for} \; \psi=\psi(x,y,z,t,u),
\end{align}
such transformation $\tilde{\psi}$ sometimes also called as one-parameter group of infinitesimal transformations. The 6-parameter version of Lie group of point transformation may be expressed as
\begin{align}
\label{KW:5}\Tilde{\psi} = \exp\left(\sum_{i=1}^{6}a_{i}X_{i}\right)\psi.
\end{align}
We know a group-invariant solution $\Psi$ corresponding to sub-group of widest invariance group can be transformed into another group invariant solution using the relation \eqref{KW:5}. The two solutions which can be connected through transformation \eqref{KW:5} shall be essentially identical. So it becomes necessary to find out only those solutions which can not be connected through this relation. The problem reduces to finding a minimal list of generators from \eqref{KW:3} which guarantees that the two group invariant solutions $\Psi_{i}$ and $\Psi_{j}$ are not connected by relation \eqref{KW:5}, that is, the problem reduces to finding an optimal list of group generators. 

Here comes the importance of adjoint transformations which will divide all the sub-groups of \eqref{KW:3} into equivalence classes, and hence, invariant solutions from such sub-groups would be essentially different. The following theorem emphasis the importance of adjoint transformations.
\begin{thm}
\normalfont 
The two group invariant solutions are essentially the same if the underlying sub-groups are adjoint or conjugate subgroups.
\begin{prof}
\normalfont Let $\Psi$ be a group invariant solution under the sub-group $H\subset G$, where $G= \{\mathrm{exp}(\epsilon X)|, X\in\mathfrak{g}, \epsilon\in\mathbb{R}\}$ is the Lie group and $H$ is its sub-group under Lie algebra $\mathfrak{h}\subset\mathfrak{g}$. Suppose $h\Psi$ be the transformation \eqref{KW:5} such that $h$ belongs to exponentiated sub-group $H$ with generator from Lie sub-algebra $\mathfrak{h}$. Since $\Psi$ is invariant under sub-group $H$, so we must have
\begin{align*}
    \Psi=h\,\Psi\quad \text{for all}\quad h\in H.
\end{align*}
Consider now a transformed solution $\tilde{\Psi}=g\,\Psi$ with $g\in G$. We may ask a simple query, that, under what sub-group $K$ the solution $\tilde{\Psi}$ would be invariant? That is, what type of sub-group $K$ would be, such that $\tilde{\Psi}= k\,\tilde{\Psi}$ for all $k\in K$? The following brief calculations answer the query
\begin{align*}
    \tilde{\Psi}=g\,\Psi=gh\,\Psi=ghg^{-1}g\Psi=ghg^{-1}\tilde{\Psi}\implies k=ghg^{-1}.
\end{align*}
This shows that, when $h\Psi$ is $H$-invariant solution and $g\Psi$ is $K$-invariant solution, then $K=K_{g}(H)=\{ghg^{-1},\, g\in G, h\in H\}$. Therefore, sub-group $K$ is the adjoint or conjugate sub-group of the group $H$ under $G$.
\end{prof}
\end{thm}
The above theorem establishes the importance of adjoint actions or adjoint transformations in partitioning the Lie algebra into equivalence classes or we can say optimal list of sub-algebras. We define adjoint transformation

\begin{align}
    \label{KW:6}\mathrm{Ad}_{\mathrm{exp}(\epsilon X_{i})}(X_{j})=\mathrm{e}^{-\epsilon X_{i}}X_{j}\mathrm{e}^{\epsilon X_{i}}=\tilde{X_{j}}(\epsilon),
\end{align}
this actually is equivalent to the operator $K_{g}(H)$. The adjoint transformation \eqref{KW:6} can be written through Lie brackets using Campbell-Hausdorff formula as
\begin{align}
\label{KW:7}\text{Ad}_{\exp(\epsilon X_{i})}\left(X_{j}\right) = X_{j}-\epsilon [X_{i},X_{j}]+\frac{\epsilon^{2}}{2}[X_{i},[X_{i},X_{j}]]-\dots,
\end{align}
where $[. , .]$ is Lie bracket defined by \autoref{KW_Table:1}. The relation \eqref{KW:7} helps to compile a table of the adjoint actions among each element in \eqref{KW:3}. All such adjoint actions are listed in the \autoref{KW_Table:2}
\begin{table}[!ht]
\centering
    \begin{tabular}{c|cccccc}
\hline\hline\\
  $\text{Ad}_{\exp(\epsilon X_{i})}\left(X_{j}\right)$ & $X_{1}$ & $X_{2}$ & $X_{3}$ & $X_{4}$ & $X_{5}$ & $X_{6}$ \\ \\\hline\hline\\
  $X_{1}$&$X_{1}$&$X_{2}$&$e^{-\epsilon}X_{3}$&$e^{-\epsilon}X_{4}$&$e^{-\epsilon}X_{5}$&$e^{-\epsilon}X_{6}$\\[1ex]
  $X_{2}$&$X_{1}$&$X_{2}$&$X_{3}+\epsilon\,X_{5}$&$X_{4}-\epsilon\,X_{6}$&$X_{5}$&$X_{6}$\\[1ex]
  $X_{3}$&$X_{1}+\epsilon\,X_{3}$&$X_{2}-\epsilon\,X_{5}+\frac{1}{2}\,\epsilon^{2}\,X_{6}$&$X_{3}$&$X_{4}$&$X_{5}-\epsilon\,X_{6}$&$X_{6}$\\[1ex]
  $X_{4}$&$X_{1}+\epsilon\,X_{4}$&$X_{2}+\epsilon\,X_{6}$&$X_{3}$&$X_{4}$&$X_{5}$&$X_{6}$\\[1ex]
  $X_{5}$&$X_{1}+\epsilon\,X_{5}$&$X_{2}$&$X_{3}+\epsilon\,X_{6}$&$X_{4}$&$X_{5}$&$X_{6}$\\[1ex]
  $X_{6}$&$X_{1}+\epsilon\,X_{6}$&$X_{2}$&$X_{3}$&$X_{4}$&$X_{5}$&$X_{6}$\\[1ex]
  \hline\hline
\end{tabular}
\caption{Table of adjoint actions defined at \eqref{KW:7}.}
\label{KW_Table:2}
\end{table}

\subsection{Construction of invariants of full adjoint action }
A real function $\phi$ defined on the Lie algebra $\mathfrak{g}$ is called \emph{an invariant} function if $\phi(\mathrm{Ad}_{g}X)=\phi(X)$ for all $X\in\mathfrak{g}$ and $g\in G$. For adjoint transformation $\mathrm{ad(X)}:\mathfrak{g}\rightarrow\mathfrak{g}$ defined by $ad(X)Y=[X,Y]$ for all $Y\in\mathfrak{g}$, the bilinear form function $K(X,Y)=trace(ad(X),ad(Y))$ is \emph{an invariant} function (for more details see the Ref. \cite{gupta2019invariant}). The special invariant function is also called Killing function and is key to classify Lie algebra into optimal list. Suppose $X=\sum_{i=1}^{6}a_{i}X_{i}$, then
\begin{align}
   \label{KW:8} {ad}(X)=\left[ \begin {array}{cccccc} 0&0&0&0&0&0\\ \noalign{\medskip}0&0&0&0
&0&0\\ \noalign{\medskip}a_{{3}}&0&-a_{{1}}&0&0&0\\ \noalign{\medskip}
a_{{4}}&0&0&-a_{{1}}&0&0\\ \noalign{\medskip}a_{{5}}&-a_{{3}}&a_{{2}}&0
&-a_{{1}}&0\\ \noalign{\medskip}a_{{6}}&a_{{4}}&a_{{5}}&-a_{{2}}&-a_{{
3}}&-a_{{1}}\end {array} \right].
\end{align}
The above adjoint transformation matrix quickly gives Killing form $K(X, X)=trace(ad(X),ad(Y))=4a_{1}^{2}$. Beside Killing form as invariant of full adjoint action, more general invariant function $\phi$ can also be calculated on the basis of procedure described in \cite{hu2015direct}. The general invariant function $\phi$ is satisfies following system of partial differential equations
\begin{eqnarray}{\label{KW:9}}
\begin{aligned}
    &\frac{\partial \phi}{\partial a_{6}}=0,\,a_{1}\,\frac{\partial \phi}{\partial a_{4}}+a_{2}\,\frac{\partial \phi}{\partial a_{6}}=0,\,a_{1}\,\frac{\partial \phi}{\partial a_{5}}+a_{3}\,\frac{\partial \phi}{\partial a_{6}}=0,\\
    &a_{3}\frac{\partial \phi}{\partial a_{5}}-a_{4}\,\frac{\partial \phi}{\partial a_{6}}=0,a_{1}\,\frac{\partial \phi}{\partial a_{3}}-a_{2}\,\frac{\partial \phi}{\partial a_{5}}-a_{5}\,\frac{\partial \phi}{\partial a_{6}}=0,\\
    &a_{3}\,\frac{\partial \phi}{\partial a_{3}}+a_{4}\,\frac{\partial \phi}{\partial a_{4}}+a_{5}\,\frac{\partial \phi}{\partial a_{5}}+a_{6}\,\frac{\partial \phi}{\partial a_{6}}=0.
\end{aligned}
\end{eqnarray}
On solving the above system of equation the general invariant function $\phi=f(a_{1}, a_{2})$ is obtained, and as we can see the Killing form is also included in this general invariant function.The actual format of this general invariant function can be predicted from the full adjoint action. 

For $X=\sum_{i=1}^{6}a_{i}X_{i}$, the repeated application of formula \eqref{KW:7} gives the following result:
\begin{align}
  \label{KW:10}\text{Ad}_{\exp(\epsilon_{5} X_{5})} \text{Ad}_{\exp(\epsilon_{6} X_{6})} \text{Ad}_{\exp(\epsilon_{1} X_{1})}\text{Ad}_{\exp(\epsilon_{3} X_{3})}\text{Ad}_{\exp(\epsilon_{4} X_{4})}\text{Ad}_{\exp(\epsilon_{2} X_{2})}(X)=\sum_{i=1}^{6}\tilde{a}_{i}X_{i},
\end{align}
where the coefficient $\tilde{a}_{i}$ are given as follows:

\begin{eqnarray}{\label{KW:11}}
\begin{aligned}
    \tilde{a}_{1}=& \,a_{1},
    \tilde{a}_{2}= \,a_{2},
    \tilde{a}_{3}=\left( a_{{1}}\epsilon_{{3}}+a_{{3}} \right) {e}^{-\epsilon_{{1}}}\\
    \tilde{a}_{4}=& \,\left( a_{{1}}\epsilon_{{4}}+a_{{4}} \right) {e}^{-\epsilon_{{1}}},
    \tilde{a}_{5}=\,a_{{1}}\epsilon_{{5}}-{e}^{-\epsilon_{{1}}}a_{{2}}\epsilon_{{3}}+{e}^{
-\epsilon_{{1}}}a_{{3}}\epsilon_{{2}}+{e}^{-\epsilon_{{1}}}a_{{5}}
,\\
    \tilde{a}_{6}=&\, a_{{1}}\epsilon_{{3}}\epsilon_{{5}}{e}^{-\epsilon_{{1}}}+a_{{1}}
\epsilon_{{6}}+\frac{1}{2}\,a_{{2}}{\epsilon_{{3}}}^{2}{e}^{-\epsilon_{{1}}}-a
_{{3}}\epsilon_{{2}}\epsilon_{{3}}{e}^{-\epsilon_{{1}}}+a_{{3}}
\epsilon_{{5}}{e}^{-\epsilon_{{1}}}+a_{{2}}\epsilon_{{4}}{e}^{-
\epsilon_{{1}}}\\
&-a_{{4}}\epsilon_{{2}}{e}^{-\epsilon_{{1}}}-a_{{5}}
\epsilon_{{3}}{e}^{-\epsilon_{{1}}}+a_{{6}}{e}^{-\epsilon_{{1}}}.
\end{aligned}
\end{eqnarray}
The first two equations in \eqref{KW:11} agree with the general invariant function $\phi=f(a_{1}, a_{2})$, that is, $a_{1}$ and $a_{2}$ are invariants of full adjoint action \eqref{KW:10}. Once the relations \eqref{KW:11} are written then it become quite easy to construct optimal list of generators. For example, $\tilde{a}_{3}$ and $\tilde{a}_{4}$ can be made zero by taking $\epsilon_{3}=-\frac{a_{3}}{a_{1}}$ and $\epsilon_{4}=-\frac{a_{4}}{a_{1}}$ respectively in \eqref{KW:11}, and this process can be repeated for further siplification of general element $X=\sum_{i=1}^{6}a_{i}X_{i}$ in \eqref{KW:3}. 
\subsection{Construction of one-dimensional optimal system}
To construct optimal, system we need to stick to the invariants $a_{1}$ and $a_{2}$. The coefficients $\tilde{a}_{i}, i=1\dots 6$ in \eqref{KW:11} can be annihilated by choosing appropriate values for $\epsilon_{i}, i=1\dots 6$. To begin with the classification process, following are the four cases depending on the values of invariants $a_{1}$ and $a_{2}$.
\begin{case}
\normalfont When $a_{1}\neq 0, a_{2}\neq 0$, we set $\epsilon_{1}=\epsilon_{2}=0$ (this means in \eqref{KW:10}, the adjoint actions $\text{Ad}_{\exp(\epsilon_{1} X_{1})}, \text{Ad}_{\exp(\epsilon_{2} X_{2})}$  are being inactivated), and on setting $\epsilon_{3}=-\frac{a_{3}}{a_{1}}$ and $\epsilon_{4}=-\frac{a_{4}}{a_{1}}$, the coefficients $\tilde{a}_{3}, \tilde{a}_{4}$ are annihilated. Further, on setting
\begin{align*}
 \epsilon_{5}=-{\frac {a_{{5}}a_{{1}}+a_{{2}}a_{{3}}}{{a_{{1}}}^{2}}}, \epsilon_{6}=-\,{\frac {2\,a_{{6}}{a_{{1}}}^{2}-2\,a_{{2}}a_{{4}}a_{{1}}+2\,a_{{
5}}a_{{3}}a_{{1}}+a_{{2}}{a_{{3}}}^{2}}{{2\,a_{{1}}}^{3}}},   
\end{align*}
the coefficients $\tilde{a}_{5}$ and $\tilde{a}_{6}$ are also annihilated. The general element $X$ finally reduce to $a_{1}X_{1}+a_{2}X_{2}$ or $X_{1}+\alpha\,X_{2}$ for $\alpha=\frac{a_{2}}{a_{1}}$.
 \end{case}
 \begin{case}
 \normalfont When $a_{1}=0, a_{2}\neq 0,$ we may take $a_{2}=1$. Setting $\epsilon_{3}=\epsilon_{4}=0$ in \eqref{KW:11}, and the selections $\epsilon_{2}=-\frac{a_{5}}{a_{3}}$ and $\epsilon_{5}=-{\frac {a_{{3}}a_{{6}}+a_{{4}}a_{{5}}}{{a_{{3}}}^{2}}}$ will annihilate the coefficients $\tilde{a}_{5}$ and $\tilde{a}_{6}$ respectively, and the general element $X$ reduce to $X_{{2}}+a_{{3}}{e}^{-\epsilon_{{1}}}X_{{3}}+a_{{4}}{e}^{-\epsilon_{{1}}}X_{{4}}$. The coefficient of $X_{3}$ can be scaled to $\pm 1$ by taking $\epsilon_{1}=\log|a_{3}|$. Final simplification shall be 
 \begin{align*}
     X=X_{2}\pm X_{3}+\beta\,X_{4},\quad \beta=\frac{a_{4}}{|a_{3}|}
 \end{align*}
 \end{case}
 \begin{subcase}
 \normalfont  When $a_{1}=0, a_{2}\neq 0, a_{3}=0,$ we may take $a_{2}=1$ in \eqref{KW:11}. We may set $\epsilon_{4}=0$, and then on taking $\epsilon_{3}=a_{5}, \epsilon_{2}=-\,{\frac {{a_{{5}}}^{2}-2\,a_{{6}}}{2a_{{4}}}},$ the coefficients $\tilde{a}_{5}$ and $\tilde{a}_{6}$ respectively, can be annihilated. So that $X$ reduce to $X_{2}+a_{{4}}{e}^{-\epsilon_{{1}}}X_{4}$. The coefficient of $X_{4}$ can be scaled to $\pm 1$ by taking $\epsilon_{1}=\log|a_{4}|$. The final simplification shall be
 \begin{align*}
     X=X_{2}\pm X_{4}
 \end{align*}
 \end{subcase}
 \begin{subcase}
 \normalfont When $a_{1}=0, a_{2}\neq 0, a_{4}=0,$ and take $a_{2}=1$ as usual. We may start by setting $\epsilon_{3}=\epsilon_{4}=0$ in \eqref{KW:11}, and on taking $\epsilon_{2}=-\frac{a_{5}}{a_{3}}$ and $\epsilon_{3}=-\frac{a_{6}}{a_{3}}$, the coefficients $\tilde{a}_{5}$ and $\tilde{a}_{6}$ respectively, can be annihilated. So that $X$ reduce to $X_{2}+a_{{3}}{e}^{-\epsilon_{{1}}}X_{3}$. The coefficient of $X_{3}$ can be scaled to $\pm 1$ by taking $\epsilon_{1}=\log|a_{3}|$. The final simplification shall be
 \begin{align*}
     X=X_{2}\pm X_{3}
 \end{align*}
 \end{subcase}
 \begin{case}
 \normalfont When $a_{1}\neq 0, a_{2}=0$, we may take $a_{1}=1$. The coefficients $\tilde{a}_{3}$ and $\tilde{a}_{4}$ can be annihilated by taking $\epsilon_{3}=-a_{3}$ and $\epsilon_{4}=-a_{4}$, such that $X$ reduce to 
 \begin{align*}
     X_{{1}}+ \left( \epsilon_{{5}}+a_{{3}}\epsilon_{{2}}{e}^{-\epsilon_{{1
}}}+a_{{5}}{e}^{-\epsilon_{{1}}} \right) X_{{5}}+ \left( \epsilon_{{6}
}+{a_{{3}}}^{2}\epsilon_{{2}}{e}^{-\epsilon_{{1}}}-a_{{4}}\epsilon_{{2
}}{e}^{-\epsilon_{{1}}}+a_{{5}}a_{{3}}{e}^{-\epsilon_{{1}}}+a_{{6}}{e}
^{-\epsilon_{{1}}} \right) X_{{6}}.
 \end{align*}
 Further, we may set $\epsilon_{1}=\epsilon_{2}=0$, and on taking $\epsilon_{5}=-a_{5}$ and $\epsilon_{6}=-a_{{5}}a_{{3}}-a_{{6}}$, the coefficients of $X_{5}$ and $X_{6}$ can be annihilated. The final simplification shall be $X=X_{1}$.
 \end{case}
 \begin{case}
 \normalfont When $a_{1}=a_{2}=0.$ The coefficients $\tilde{a}_{5}$ and $\tilde{a}_{6}$ can be annihilated by taking $\epsilon_{2}=-\frac{a_{5}}{a_{3}}$ and $\epsilon_{5}=-{\frac {a_{{3}}a_{{6}}+a_{{4}}a_{{5}}}{{a_{{3}}}^{2}}}$ respectively. The final simplification shall be $X=a_{3}X_{3}+a_{4}X_{4}=X_{3}+\gamma\,X_{4}$ for $\gamma=\frac{a_{4}}{a_{3}}$.
 \end{case}
 
 \begin{subcase}
 \normalfont When $a_{1}=a_{2}=a_{3}=0.$ On taking $\epsilon_{2}=0$ and $\epsilon_{3}=\frac{a_{6}}{a_{5}}$, the final simplification shall be $X=X_{4}+\delta\,X_{5}$ for $\delta=\frac{a_{5}}{a_{4}}$.
 \end{subcase}
 \begin{subcase}
 \normalfont  When $a_{1}=a_{2}=a_{4}=0.$ On taking $\epsilon_{2}=-\frac{a_{5}}{a_{3}}$ and $\epsilon_{5}=-\frac{a_{6}}{a_{3}}$, the final simplification shall be $X=X_{3}$.
 \end{subcase}
 \noindent We then have one-dimensional optimal system $\Theta_{1}$ as follows:
 \begin{eqnarray}{\label{KW:12}}
\begin{aligned}
   & X_{1}+\alpha\,X_{2}, X_{2}\pm X_{3}+\beta\,X_{4},\\
    &X_{2}\pm X_{4},X_{2}\pm X_{3}, X_{1},\\
    &X_{3}+\gamma\,X_{4}, X_{4}+\delta\,X_{5}, X_{3}.
\end{aligned}
\end{eqnarray}
\subsection{Construction of two-dimensional optimal system}
To construct two-dimensional optimal system $\Theta_{2}$, we follow the standard procedure given in \cite{ovsi}. To prepare list of sub-algebras for $\Theta_{2}$, we need to find sub-algebra of the type $\mathscr{H}\left(x_{i}, x_{j}\right)$, where the element $x_{i}$ is taken from the list of sub-algebras $\Theta_{1}$ obtained at \eqref{KW:12}, and $x_{j}$ belongs to the normalizer sub-algebra $\text{Nor}_{\mathfrak{g}}\left(x_{i}\right)=\left\{x\in \mathfrak{g} | \left[x, x_{i}\right]\in x_{i}\right\}$, that is, $\left[x_{i}, x_{j}\right]= \lambda x_{i}$. While selecting $x_{j}$, one can avoid occurrence of $x_{i}$ in $x_{j}$ by selecting $x_{j}$ from factor algebra $\text{Nor}_{\mathfrak{g}}\left(x_{i}\right)/x_{i}$, where  the normalizer $\text{Nor}_{\mathfrak{g}}\left(x_{i}\right)$ can be obtained by setting
\begin{align}
\label{KW:13}\left[x_{i}, \sum_{j=1}^{6} a_{j}X_{j}\right]=\lambda x_{i},
\end{align}
where $[\cdot\, , \cdot]$ is the usual Lie bracket and $\lambda$ is an arbitrary constant. In the relation \eqref{KW:13}, the coefficients of $X_{i}$  can equated to find out all possible non-zero $a_{i}$' for the construction of $\text{Nor}_{\mathfrak{g}}\left(x_{i}\right)$ and hence $\text{Nor}_{\mathfrak{g}}\left(x_{i}\right)/x_{i}$. So the following list of two-dimensional sub-algebras is obtained:
 \begin{eqnarray}{\label{KW:14}}
\begin{aligned}
   &\mathscr{H}_{1}(X_{2}\pm X_{3}+\beta\,X_{4},a\,X_{5}+b\,X_{6}), \mathscr{H}_{2}(X_{2}\pm X_{4},a\,X_{5}+b\,X_{6}),\\
  & \mathscr{H}_{3}(X_{2}\pm X_{3},-a\,X_{4}+a\,X_{5}+b\,X_{6}),\mathscr{H}_{4}(X_{1},X_{2}),\\
 & \mathscr{H}_{5}(X_{3}+\gamma\,X_{4}, X_{6}),\mathscr{H}_{6}(X_{4}+\delta\,X_{5},-a\delta\,X_{2}+a\,X_{3}+b\,X_{6}),\\
 &\mathscr{H}_{7}(X_{3},a\,X_{1}+b\,X_{4}+c\,X_{6}).
\end{aligned}
\end{eqnarray}
In following, we shall try to simplify each pair in the list \eqref{KW:14} with the adjoint action \eqref{KW:7} as much as possible. The two elements $\{x_{1}, x_{2}\}$ and $\{x_{1}^{\prime}, x_{2}^{\prime}\}$ equivalent under adjoint action if
\begin{eqnarray}{\label{KW:15}}
\begin{aligned}
x_{1}^{\prime}=&\,k_{1}\,\text{Ad}_{\exp(\epsilon X)}\left(x_{1}\right)+k_{2}\,\text{Ad}_{\exp(\epsilon X)}\left(x_{2}\right),\\
x_{2}^{\prime}=&\,k_{3}\,\text{Ad}_{\exp(\epsilon X)}\left(x_{1}\right)+k_{4}\,\text{Ad}_{\exp(\epsilon X)}\left(x_{2}\right),
\end{aligned}
\end{eqnarray}
where constants $k_{i}$ are such that at least one of the pair $(k_{1}, k_{4})$ or $(k_{2}, k_{3})$ is non-zero and $X$ is general element of Lie algebra $\mathfrak{g}$ given at \eqref{KW:3}. The following inverse version of \eqref{KW:15} is more appropriate for classification of 2-dimensional subalgebra.
\begin{align}{\label{KW:16}}
\text{Ad}_{\exp(\epsilon X)}\left(x_{1}\right)=k_{1}\,x_{1}^{\prime}+k_{2}\,x_{2}^{\prime}, \,
\text{Ad}_{\exp(\epsilon X)}\left(x_{2}\right)=k_{3}\,x_{1}^{\prime}+k_{4}\,x_{2}^{\prime}.
\end{align}
As an example, we consider $\mathscr{H}_{7}(X_{3},a\,X_{1}+b\,X_{4}+c\,X_{6})$ for simplification under adjoint actions. We must consider different possible values of the triplet $(a,b,c)$, For example, $(a,b,c)$; all nonzero constants, $(a,b,0), (a,0,c), (0,b,c)$; one constant is zero, $(a,0,0), (0,b,0), (0,0,c)$; two constants are zero.

 For $(a,b,c)$; all nonzero constants. Taking full adjoint action \eqref{KW:10} on $X_{3}$ and $a\,X_{1}+b\,X_{4}+c\,X_{6}$, the equations \eqref{KW:16} can be written as follow:
\begin{subequations}{\label{KW:17}}
\begin{align}
\label{KW:17a}&{e}^{-\epsilon_{{1}}} \left(-X_{{6}}\epsilon_{{2}}\epsilon_{{3}}+
\epsilon_{{2}}X_{{5}}+X_{{6}}\epsilon_{{5}}+X_{{3}} \right) =k_{{1}}X_
{{3}}+k_{{2}} \left( a^{\prime}X_{{1}}+b^{\prime}X_{{4}}+c^{\prime}X_{{6}} \right),\\
&aX_{{1}}+a\epsilon_{{3}}{e}^{-\epsilon_{{1}}}X_{{3}}+ \left( a\epsilon
_{{4}}{e}^{-\epsilon_{{1}}}+b{e}^{-\epsilon_{{1}}} \right) X_{{4}}\nonumber\\
\label{KW:17b}&+a
\epsilon_{{5}}X_{{5}}+ \left( a\epsilon_{{3}}\epsilon_{{5}}{e}^{-
\epsilon_{{1}}}+a\epsilon_{{6}}-b\epsilon_{{2}}{e}^{-\epsilon_{{1}}}+c
{e}^{-\epsilon_{{1}}} \right) X_{{6}}=k_{{3}}X_{{3}}+k_{{4}} \left( a^{\prime}X_{{1}}+b^{\prime}X_{{4}}+c^{\prime}X_{{6}}\right).
\end{align}
\end{subequations}
The coefficients of $X_{1}, X_{4}$ give $a^{\prime}=0$ and $b^{\prime}=0$ respectively. So the final simplification of $\mathscr{H}_{7}$ shall be $\mathscr{H}_{7}(X_{3},X_{6})$, and same simplification for other cases too. Repeating this procedure for other 2-dimensional sub-algebra in \eqref{KW:14}, the final simplification is given as follow:
\begin{eqnarray}{\label{KW:18}}
\begin{aligned}
   \mathscr{G}_{1}(X_{2}\pm X_{3}+\beta\,X_{4},X_{5}),\mathscr{G}_{2}(X_{2}\pm X_{3}+\beta\,X_{4},X_{6}),\\
   \mathscr{G}_{3}(X_{2}\pm X_{4},\,X_{5}),\mathscr{G}_{4}(X_{2}\pm X_{4},\,X_{6}), \mathscr{G}_{5}(X_{2}\pm X_{3}, X_{6}),\\
   \mathscr{G}_{6}(X_{1}, X_{2}), \mathscr{G}_{7}(X_{3}+\gamma\,X_{4}, X_{6}), \mathscr{H}_{9}(X_{4}+\delta\,X_{5},X_{6}), \mathscr{H}_{8}(X_{3},X_{6}).
\end{aligned}
\end{eqnarray}
In similar fashion, a three-dimensional optimal system can also be constructed, we have avoided writing that here due to space constraints. In following section, we show reductions with respect optimal systems \eqref{KW:14} and \eqref{KW:18}.
\begin{rem}
\normalfont We may have continued for symmetry reductions of the equation \eqref{KW:1} but that would not fruitful as the equation is of the eighth order. Its order can be reduced by one or by two using one-dimensional sub-algebra \eqref{KW:12} and two-dimensional sub-algebra \eqref{KW:14} respectively, and even then the equation shall remain of a higher order. And unfortunately, the symmetry reductions of \eqref{KW:1} can not be accomplished with one and two-dimensional Lie algebras \eqref{KW:12} and \eqref{KW:18} respectively, until the arbitrary functions are set to polynomials. 
\end{rem}
\section{Painlev$\acute{\rm e}$ analysis }
In remark \ref{KW_RM:1}, the non-existence of Virasoro algebra suggests that the Kac-Wakimoto equation may not be integrable, and with the Painlev$\acute{\rm e}$ analysis we try to strengthen this claim. The Painlev$\acute{\rm e}$ analysis is a well-established tool for investigating complete integrability of nonlinear partial differential equations. Looking at the singularity structure of the equation one can predict the complete integrability, for details the refs. \cite{weiss,ptest3,conte2008painleve} can be seen. In this section,  we investigate the singularity structure of equation \eqref{KW:1} based on the algorithm suggested by Weiss-Tabor-Carnevale. 
The Laurent expansion can be taken in the {following} form
\begin{align}{\label{KW:19}}
u = \phi^{\alpha}\sum_{j = 0}^{\infty}u_{j}\,\phi^{j},\quad \text{where } \phi  = \phi(x,y,z,t).
\end{align}
When \eqref{KW:19} is substituted into \eqref{KW:1}, the following conditions need to be satisfied for valid Painlev$\acute{\rm e}$ property.
\begin{enumerate}
    \item {$\alpha$ to be an integer.}
    \item {$\phi$ to be analytic function of $(x,y,z,t)$.}
\item {The equations for $u_{j}$ to have self-consistent solutions.}
\end{enumerate}
The quick survey of leading order analysis gives
\begin{align*}
&\text{first branch:}\quad\alpha = \,-1,\;u_{0} = 2\,\phi_{x},\\
&\text{second branch:}\quad\alpha = \,-1,\;u_{0} = 4\,\phi_{x},\\
&\text{third branch:}\quad\alpha = \,-1,\;u_{0} = 6\,\phi_{x}.
\end{align*}
The resonant points can be determined by substituting
\begin{align}
u = u_{0}\,\phi^{-1}+u_{j}\,\phi^{j-1},
\end{align}
into \eqref{KW:1} and on retaining most singular part, the resonance points at each branch obtained as follows:
\begin{align*}
&\text{first branch:}\quad j=-1, 1, 2, 3, 4, 5, 8 ,14,\\
&\text{second branch:}\quad j=-2, -1, 1, 2, 3, 8, \frac{25+\sqrt{65}}{2}, \frac{25-\sqrt{65}}{2},\\
&\text{third branch:}\quad j=-3, -2, -1, 1, 8, 10, \frac{23+\sqrt{193}}{2}, \frac{23-\sqrt{193}}{2}.
\end{align*}
The non-integral resonance points at the second and third branches suggest that the Kac-Wakimoto equation fails to pass the Painlev$\acute{\rm e}$ property, and therefore can not be integrable.
\section{Conclusion}
The eight-order (3+1)-dimensional Kac-Wakimoto equation is studied for its integrability. This equation has already shown to be non-integrable using Hirota's bilinear method as it does not have three-soliton solutions \cite{pekcan2016kac}. In this paper, the non-integrability of the equation is checked from different angles, such as; the infinite-dimensional Lie algebra for Kac-Wakimoto equation does not have Virasoro like structure, which in fact common characteristic of the most of integrable equations, and absence of Virasoro like structure implies that the equation may be non-integrable. In order to further strengthen the argument for the non-integrability of equation, the Painlev$\acute{\rm e}$ property is also checked, which also turned out to be negative. 

It is important to note that there are some integrable systems that are invariant under finite or infinite-dimensional Lie algebra without having Virasoro like structure (see ref. \cite{senthil1998lie}), but all non-integrable system that are invariant finite or infinite-dimensional Lie algebra does not have Virasoro like structure. In this way, the presence  of Virasoro like structure can be a week predictor of integrability and it could be used along with other strong predictors of integrability.

\end{document}